# Pseudogap, Fermi arc, and Peierls-insulating phase induced by 3D-2D crossover in monolayer VSe$_2$

Yuki Umemoto[1], Katsuaki Sugawara*[1,2,3], Yuki Nakata[1], Takashi Takahashi[1,2,3], and Takafumi Sato[1,3]

[1]*Department of Physics, Tohoku University, Sendai 980-8578, Japan,*

[2]*WPI-Advanced Institute for Materials Research, Tohoku University, Sendai 980-8577, Japan*

[3]*Center for Spintronics Research Network, Tohoku University, Sendai 980-8577, Japan*

## ABSTRACT

One of important challenges in condensed-matter physics is to realize new quantum states of matter by manipulating the dimensionality of materials, as represented by the discovery of high-temperature superconductivity in atomic-layer pnictides and room-temperature quantum Hall effect in graphene. Transition-metal dichalcogenides (TMDs) provide a fertile platform for exploring novel quantum phenomena accompanied by the dimensionality change, since they exhibit a variety of electronic/magnetic states owing to quantum confinement. Here we report an anomalous metal-insulator transition induced by 3D-2D crossover in monolayer 1$T$-VSe$_2$ grown on bilayer graphene. We observed a complete insulating state with a finite energy gap on the entire Fermi surface in monolayer 1$T$-VSe$_2$ at low temperatures, in sharp contrast to metallic nature of bulk. More surprisingly, monolayer 1$T$-VSe$_2$ exhibits a pseudogap with Fermi arc at temperatures above the charge-density-wave temperature, showing a close resemblance to high-temperature cuprates. This similarity suggests a common underlying physics between two apparently different systems, pointing to the importance of charge/spin fluctuations to create the novel electronic states, such as pseudogap and Fermi arc, in these materials.

## 1. INTRODUCTION

Atomic-layer transition-metal dichalcogenides (TMDs) MX$_2$ (M = transition metal, X = chalcogen) are recently attracting a particular attention since they exhibit a variety of physical properties such as spin- and valley-coupled 2D superconductivity (Ising superconductivity) [1,2], exciton Hall effect [3], and exciton-driven charge-density wave (CDW) [4,5]. Atomic-layer TMDs are expected to provide a useful platform to realize nano electronic devices such as biosensors [6], rechargeable batteries [7], and optical devices [8].

Amongst TMDs, bulk 1$T$-VSe$_2$ is known to be a typical CDW material showing an in-plane commensurate (4×4) and an out-of-plane incommensurate CDW below the transition temperature ($T_{CDW}$) of 110 K [9,10]. The CDW mechanism is well understood in the framework of conventional Fermi-surface (FS) nesting, where the nesting vector connects a part of warped 3D electron pocket centered at the L point (called partial nesting) [11-13]. Recently it was reported that exfoliation of bulk crystal leads to an intriguing change in $T_{CDW}$; i.e. $T_{CDW}$ gradually decreases with reducing the number of layers [14], but suddenly exhibits a characteristic upturn at around the film thickness of 10 nm, reaching ~130 K much higher than bulk $T_{CDW}$ (110 K) [15]. Further, very recently, monolayer VSe$_2$ was fabricated and the low energy electron diffraction (LEED), angle-resolved photoemission spectroscopy (ARPES) and scanning tunneling microscopy (STM) have revealed that the $T_{CDW}$ is enhanced in monolayer (~140 K) compared to bulk [16,17,18]. While these experimental results suggest a possible connection between the stabilization of CDW and the quantum size effect, the momentum dependence of the CDW gap and its temperature evolution are still unclear. It is thus of particular importance to elucidate the detailed electronic states of monolayer VSe$_2$ and unveil the key ingredients relevant to the characteristic CDW properties.

In this paper, we report ARPES study on monolayer 1$T$-VSe$_2$ film grown epitaxially on bilayer graphene on SiC(0001). We found (i) a small circular and a large triangular hole pocket centered at the Γ and K points, respectively, (ii) an insulating energy gap on the entire Fermi surface below $T_{CDW}$, and (iii) a pseudogap with a Fermi arc above $T_{CDW}$. The present ARPES results suggest the importance of quantum fluctuations to the CDW transition in monolayer 1$T$-VSe$_2$.

## 2. RESULTS AND DISCUSSION

First we explain the fabrication of high-quality single-crystalline monolayer VSe$_2$. We used the van-der-Waals molecular-beam-epitaxy (MBE) technique [19] with bilayer graphene grown on silicon carbide (SiC) as a substrate [4,18,19]. Figure 1(b) shows the reflection high-energy electron diffraction (RHEED) pattern of pristine bilayer graphene grown on 6$H$-SiC(0001). We clearly observe the 1×1 streak pattern and the 6√3×6√3$R$30° spots originating from bilayer graphene and underlying buffer structure, respectively [4,20,21]. After evaporating V atoms in Se atmosphere onto the substrate kept at 400 ºC, the RHEED intensity due to graphene and buffer structure is reduced, and instead a new sharp 1×1 streak pattern appears (Fig. 1(c)) similarly to the case of other monolayer TMD films on bilayer graphene [4,20,21], indicating the formation of monolayer VSe$_2$ on bilayer graphene. As seen in the *ex-situ* AFM image (Fig. 1(d)), several monolayer islands are recognized on the bilayer graphene substrate.

Next, we determined the electronic structure of monolayer VSe$_2$ by *in-situ* ARPES. Figure 2(a) shows the ARPES-intensity plot in the valence-band region for monolayer VSe$_2$ measured at room temperature along the K-Γ-M cut in the Brillouin-zone (BZ). It is noted that substrate graphene does not have the energy bands (both π and σ bands) in this energy and momentum region, so that the energy bands in Fig. 2(a) are attributed to VSe$_2$. Figure 2(b) shows the calculated band structure for free-standing monolayer 1*T*-VSe$_2$. One can see in Fig. 2(a) several dispersive bands such as a highly dispersive holelike Se 4*p* band and a less-dispersive V 3*d* band near the Fermi level ($E_F$). As seen in Fig. 2(c), the energy position of these bands shows no discernible photon-energy ($h\nu$) dependence, supporting the 2D nature of electronic structure. The good agreement in the overall band structure between the experiment and the calculation, as seen in Fig. 2(a) and (b), indicates that the fabricated VSe$_2$ monolayer certainly forms the 1*T* structure. The lattice constant estimated from the wave vector at the M point is ~3.35 Å, which is almost the same as that of bulk (~3.36 Å), suggesting that the VSe$_2$ monolayer film fabricated on bilayer graphene is free from epitaxial strain and regarded to be nearly free-standing.

To see more clearly the energy dispersion of the metallic V 3*d* band responsible for the anomalous physical properties, we show in Fig. 2(d) the near-$E_F$ ARPES-intensity plot at room temperature measured along the K-Γ-M cut. To visualize the band dispersion above $E_F$, the ARPES intensity was divided by the Fermi-Dirac distribution (FD) function. The electronic state around the Γ point seems to consist of two different holelike branches originating from the Se 4*p* and V 3*d* orbitals (see Fig. 2(d)). The Se 4*p* band shows a steep hole-like dispersion with a top at ~70 meV in binding energy ($E_B$) *below* $E_F$ at the Γ point, meaning that the Se 4*p* state does not contribute to the FS. The V 3*d* band, on the other hand, has a local maximum at $E_B$ ~ -30 meV *above* $E_F$ at the Γ point, disperses toward higher $E_B$ along the Γ-M cut, and reaches the bottom of $E_B \sim 400$ meV at the M point. In the direction of Γ-K cut, this V 3$d$ band crosses $E_F$ near the Γ point as in the Γ-M cut, but immediately disperses upward and crosses $E_F$ again. This anomalous dispersive behavior of the V 3$d$ band is distinctly different from that of bulk VSe$_2$ [11-13], suggesting that the change in the dimensionality and the resultant absence of interlayer interaction strongly influence the electronic states of monolayer VSe$_2$.

To investigate how the reduction of dimensionality modifies the FS topology, we have performed ARPES measurements in the vicinity of $E_F$. Figure 2(e) shows the ARPES intensity at $E_F$ plotted as a function of 2D wave vector. One can recognize a small circular hole pocket (α) centered at the Γ point and a large triangular hole pocket (β) at the K point, which is consistent with the previous result [16]. Interestingly, the shape of β pocket is markedly different from that of bulk at the $k_z=\pi$ plane (A-H-L plane), while it looks similar to that at $k_z=0$ (Γ-K-M plane) [11-13]. The bulk FS at $k_z=\pi$ is an ellipsoidal electron pocket centered at the L point (corresponding to the $\overline{\text{M}}$ point in the surface BZ), while that at $k_z=0$ has a triangular shape centered at the K point [11-13]. The α pocket of monolayer has no counterpart in bulk in both experiment and calculation. This difference between monolayer and bulk is not attributed to the carrier-doping effect, because the total carrier number of monolayer VSe$_2$ estimated from the FS volume is ~1 e-/unit, indicating that the monolayer sample keeps stoichiometry. Therefore, we conclude that the observed anomalous FS topology is an intrinsic characteristic of monolayer 1$T$-VSe$_2$.

Having established the FS topology, a next important question is on the CDW properties of monolayer. To address this question, we have performed temperature-dependent ARPES measurements near $E_F$, and show in Fig. 3(a) and (b) the temperature dependence of energy distribution curves (EDCs) at two representative Fermi vectors ($k_F$'s) on the α and β pockets (points A and B in the inset to Fig. 3(a) and (b), respectively). At point A, one can clearly see a well-defined peak near $E_F$ in the EDC, which is gradually sharpened on decreasing temperature. At $T = 40$ K, the leading-edge midpoint is apparently shifted from $E_F$ toward higher $E_B$ and as a result the spectral weight at $E_F$ is significantly suppressed. This is a clear evidence for energy-gap opening and is better visualized in the symmetrized EDCs in Fig. 3(c) in which one can recognize a two-peaked structure persisting up to $T = 120$ K. Since the two-peaked structure disappears above 140 K, it is inferred that the gap starts to open at around this temperature. It is remarked that the observed $T_{CDW}$ (140 K) of monolayer $VSe_2$ is in agreement with the recent study on a MBE-grown monolayer film [16], but is apparently higher than those of both bulk (~ 110 K) and exfoliated multilayer thin films (< 130 K) [14,15]

As shown in Fig. 3(b), the EDC of β pocket (point B) at $T = 40$ K exhibits a leading-edge shift similarly to the α pocket, whereas the overall spectral feature is much broader and a pseudogap-like structure with an energy scale of ~200 meV is observed as indicated by a black circle in Fig. 3(b). On increasing temperature, the leading-edge midpoint gradually shifts toward $E_F$, showing a gradual gap-closure with temperature. As shown in the symmetrized EDCs in Fig. 3(d), one can see suppression of the spectral weight around $E_F$ even at room temperature, in sharp contrast to a single-peaked feature of the α pocket at high temperature (Fig. 3(c)). This suggests that the pseudogap on the β pocket survives even at room temperature. To evaluate the gap-opening behavior in more detail, we plot in Fig. 3(e) the temperature dependence of the energy position of the gap magnitude $\Delta_{gap}$ estimated from the peak position in the symmetrized EDCs at point A on the α pocket. Upon decreasing temperature, $\Delta_{gap}$ exhibits a sudden jump at ~140 K and reaches ~70 meV at $T = 40$ K. This confirms that the onset temperature of gap opening is ~ 140 K. As seen in Fig. 3(f), the $\Delta_{gap}$ value at $T = 40$ K at point B (~ 230 meV) is about 3 times larger than that of point A, and surprisingly keeps 170-180 meV even above 140 K, indicating the existence of a pseudogap at high temperature.

To see the *k*-dependence of gap, we plot in Fig. 4(b) the symmetrized EDCs at $T$ = 40 K measured at various $k_F$'s on the α and β pockets (see Fig. 4(a)). In both α and β pockets, we observe a two-peaked structure in the symmetrized EDC anywhere on the FS at 40 K, indicating that an energy gap opens on the entire FS. This strongly suggests the insulating nature of monolayer VSe$_2$ at $T$ = 40 K in contrast to the metallic nature of bulk VSe$_2$ [11,12]. This means that the metal-insulator transition is induced by the 3D-2D crossover in 1*T*-VSe$_2$. It is also noted that the EDC on the α pocket is always sharp while that on the β pocket is broad with a hump structure. Intriguingly, the energy scale characterizing the hump structure (same as $\Delta_{gap}$ and indicated by filled circles in Fig. 4(b) and (c)) shows a finite *k*-dependence; it is ~ 200 meV at points 6 and 7, gradually decreasing on approaching the Γ point, and reaches ~ 100 meV at point 3 on the corner of the triangular pocket. As shown in Fig. 4(d), the $\Delta_{gap}$ of α pocket at $T$ = 40 K is almost constant against the variation of the FS angle $\theta$ (defined in Fig. 4(a)), whereas that of the β pocket (Fig. 4(e)) is highly anisotropic with the minimum (~ 100 meV) at $\theta$ ~ 0° and the maximum (~ 200 meV) at $\theta$ ~ ± 60°. As shown in Fig. 4(c), the symmetrized EDC mesured at room temperature at point 7 still shows a two-peaked structure due to the pseudogap opening. The gap value gradually decreases on moving toward the corner of triangular FS (from point 6 to 5), and completely closes at points 4 and 3. Since the gap vanishes not at a single *k* point but in a finite *k* region, there exists a *Fermi arc* at room temperature. This feature is better visualized in Fig. 4(e) where the zero-gap region extends in a finite $\theta$ range ($|\theta|$ < 10°). Thus, the present experimental result strongly suggests the coexistence of pseudogap and Fermi arc on a single FS, showing a close resemblance to high-temperature cuprate superconductors [22,23].

The present result has an important implication on the interplay between the dimensionality and the Peierls transition. We found that monolayer VSe$_2$ shows the insulating ground state in contrast to the metallic nature of bulk. Previous ARPES studies suggested that the CDW in bulk 1$T$-VSe$_2$ is triggered by the FS nesting [11,12]. As shown in Fig. 4(f), the in-plane nesting vector (1/4, 1/4) in bulk 1$T$-VSe$_2$ connects a part of the elongated pocket centered at the L point, leaving other poorly nested region un-gapped, and thereby keeps a metallic nature even below $T_{CDW}$. A recent theoretical calculation reports that the FS topology of monolayer resembles that of bulk at $k_z = \pi$ (Fig. 4(f)) and the CDW in monolayer is likely triggered by the in-plane nesting identical to that of bulk [24]. However, this is not the case because, as found in the present experiment, the FS topology of monolayer is totally different from that of bulk at $k_z = \pi$. The fully gapped insulating behavior in monolayer should be attributed to a strong nesting owing to the triangular shape with a long straight segment in the β pocket (see Fig. 4(g)). This almost perfect nesting would strongly stabilize the CDW state, enhancing the $T_{CDW}$, and leads to the gap opening even on the α pocket, which supports the observation of the (4x4) superstructure in the LEED pattern of monolayer VSe$_2$, similarly to that of bulk [16]. In fact, the estimated maximal $2\Delta_{Gap}/k_B T_{CDW}$ value for monolayer VSe$_2$ (~38 for the β pocket) is much larger than that of bulk (~20), supporting the more stable CDW state in monolayer.

Perhaps the present result is the first realization of an atomic-layer Peierls insulator driven by the 3D-2D crossover. The present discovery of Peierls insulating ground state in monolayer VSe$_2$ supports the recent report that bulk 1$T$-VSe$_2$ becomes to show an insulating behavior below $T_{CDW}$ in transport measurement when a SnSe/PbSe layer is inserted (intercalated) between adjacent VSe$_2$ monolayers [25,26]. This drastic change in the transport properties could be understood in terms of the enhancement of 2D nature and the Peierls transition in intercalated bulk VSe$_2$ because the intercalated layer would suppress the interlayer interaction between adjacent VSe$_2$ sheets and as a result each VSe$_2$ layer separated by intercalated layer would behave as like a free-standing monolayer.

The present result also sheds lights on the long-standing debate on the origin of pseudogap in other systems such as cuprates. In bulk 1$T$-VSe$_2$, a pseudogap with a finite Fermi-edge cut-off was observed below $T_{CDW}$ [11] and has been attributed to the $k_z$-selective partial nesting owing to the large $k_z$ broadening. However, this scenario is apparently not applicable to monolayer because $k_z$ is not a good quantum number and the $k_z$-selective nesting does not occur in monolayer. A plausible scenario to explain the appearance of pseudogap is the spin fluctuations. It is worthwhile to point out a recent discovery of ferromagnetism at room temperature in monolayer VSe$_2$ grown on HOPG or MoS$_2$ [27]. If the strong ferromagnetic fluctuations [28] promote the inelastic electron scattering on the β pocket, the pseudogap may arise in a wide temperature range. However, at this stage, it is too early to speculate in this line, since the present and previous ARPES studies [16, 17] commonly show no clear evidence for the exchange splitting of the V-3$d$ band as predicted in the theory [29]. This may be because of the considerably small splitting or a possible substrate dependence on the occurrence of ferromagnetism (bilayer graphene vs. HOPG or MoS$_2$). A spin-resolved ARPES and/or x-ray magnetic circular dichroism study for monolayer and multilayer films would help to resolve this problem.

The pseudogap may be explained in the framework of the strong charge (CDW) fluctuations around the well-nested segments of triangular FS (possible nesting vector is depicted in Fig. 4(g)) [30]. In this case, the corner of triangular FS would be relatively poorly nested, resulting in the suppression of CDW fluctuations and the resultant recovery of a conventional metallic state characterized by the Fermi arc. This indicates a close similarity to cuprates superconductors where the pseudogap and Fermi arc are seen in the *k* region with flat FS segments spanned by the nesting vector and the nodal region with a poor nesting condition, respectively [21,31,32]. The present result suggests common underlying physics for monolayer VSe$_2$ and cuprate superconductors.

## 3. CONCLUSION

We have performed an ARPES study on monolayer 1*T*-VSe$_2$ grown on bilayer graphene, and experimentally determined the band structure and FS. We found a small circular and a large triangular hole pocket centered at the Γ and K points, respectively. The triangular pocket likely drives the Peierls transition due to the almost perfect nesting, leading to the insulating ground state characterized by an energy-gap opening on the entire FS. We also found a pseudogap with a Fermi arc at above the CDW temperature, similarly to high-temperature cuprates. The present result opens a pathway toward understanding the interplay between the dimensionality and the novel physical properties in atomic-layer TMDs.

## 4. METHODS

Monolayer VSe$_2$ film was fabricated by the MBE method [19]. Bilayer graphene, used as a substrate, was prepared by annealing an *n*-type Si-rich 6*H*-SiC(0001) single-crystal wafer by resistive heating at 1100ºC for 20 min in an ultrahigh vacuum better than $1.0 \times 10^{-9}$ Torr. Monolayer VSe$_2$ film was grown by co-evaporating V and Se atoms on a bilayer graphene substrate kept at 400ºC. The as-grown film was annealed at 350ºC for 30 min, and transferred to the ARPES measurement chamber without breaking vacuum. The film thickness was monitored by a quartz-oscillator thickness meter (Inficon) and AFM (Park NX10). ARPES measurements were carried out using a MBS-A1 electron-energy analyzer with a high-flux helium discharge lamp and a toroidal grating monochromator in Tohoku University and a Omicron-Scienta SES2002 electron-energy analyzer with synchrotron radiation in Photon Factory (KEK). The energy and angular resolutions were set at 16-30 meV and 0.2º, respectively. The Fermi level of samples was referenced to that of a gold film deposited onto the sample substrate. To perform $h\nu$-dependent ARPES measurements in KEK by avoiding contamination of the sample surface during the transfer, we covered the VSe$_2$ film with amorphous Se immediately after the epitaxy at Tohoku University, and de-capped it by annealing the film under ultrahigh vacuum at KEK. First-principles band-structure calculations for free-standing monolayer VSe$_2$ was carried out by using the Quantum Espresso code [33] with generalized gradient approximation [34]. The plane-wave cutoff energy and the $k$-point mesh were set to be 30 Ry and 12×12×1, respectively. The crystal structure was relaxed using the supercell geometry with a vacuum region of more than 10 Å.

ACKNOWLEDGMENTS

We thank I. Watanabe, H. Oinuma, D. Takane, S. Souma, K. Horiba, and H. Kumigashira for their assistance in ARPES experiments. This work was supported by JSPS KAKENHI Grants (JP25107003, JP15H05853, JP15H02105, JP17H01139), KEK-PF (Proposal No, 2015S2-003, 2018S2-001), Grant for Basic Science Research Projects from the Sumitomo Foundation, Science Research Projects from Iketani Science and Technology Foundation, the Program for Key Interdisciplinary Research, and World Premier International Research Center, Advanced Institute for Materials Research. Y.N. acknowledges support from GP-Spin at Tohoku University.

AUTHOR CONTRIBUTIONS

Y.U., K.S. and Y.N. carried out the fabrications of thin films, their characterization and ARPES measurements. Y.U., K.S., T.T. and T.S. finalized the manuscript with input from all the authors.

**Competing Interests**

We have no competing financial and non-financial interests.

**FIGURE LEGENDS**

**Figure 1** (a) Crystal structure of monolayer $VSe_2$ with trigonal prismatic (1$T$) structure. Left panel shows the unit cell of monolayer $VSe_2$. Right panel shows the top and side views of crystal. (b) and (c) RHEED patterns of bilayer (BL) graphene and monolayer $VSe_2$, respectively, obtained along the [1$\bar{1}$00] direction of SiC(0001) substrate. (d) Typical AFM image of monolayer $VSe_2$ islands on bilayer graphene. The trigonal and/or hexagonal shape of islands supports the single-crystalline growth of $VSe_2$.

**Figure 2** (a) Valence-band ARPES intensity of monolayer $VSe_2$ measured along the K-Γ-M cut at room temperature with the He-Iα resonance line ($hv$ = 21.218 eV). (b) Calculated band structure obtained from the first-principles band-structure calculations for free-standing monolayer 1$T$-$VSe_2$. (c) Photon-energy dependence of ARPES spectrum at the Γ point. (d) Near-$E_F$ ARPES-intensity plot along the K-Γ-M cut divided by the FD function. (e) ARPES intensity at $E_F$ plotted as a function of 2D wave vector. Intensity at $E_F$ was obtained by integrating the intensity within ± 10 meV with respect to $E_F$, and is folded by taking into account the six-fold symmetry. Filled red circles correspond to the position of $k_F$'s.

**Figure 3** (a) and (b) Temperature dependence of EDC at points A and B on the FS, respectively. (c) and (d) Same as (a) and (b), but symmetrized with respect to $E_F$. (e) and (f) Temperature dependence of the gap magnitude ($\Delta_{gap}$) at points A and B, respectively.

**Figure 4** (a) Schematic FS and $k_F$ points where the EDCs in (b) and (c) were measured. Definition of FS angle $\theta$ is also shown. (b) and (c) Symmetrized EDCs near $E_F$ of monolayer VSe$_2$ at $T = 40$ K and room temperature, respectively, measured at various $k_F$ points shown in (a). (d) and (e) $\theta$-dependence of $\Delta_{gap}$ for $\alpha$ and $\beta$ pockets, respectively, at $T = 40$ K and room temperature. (f) and (g) Schematic view of FS topology and possible nesting vectors (solid purple arrows) in bulk and monolayer VSe$_2$, respectively.

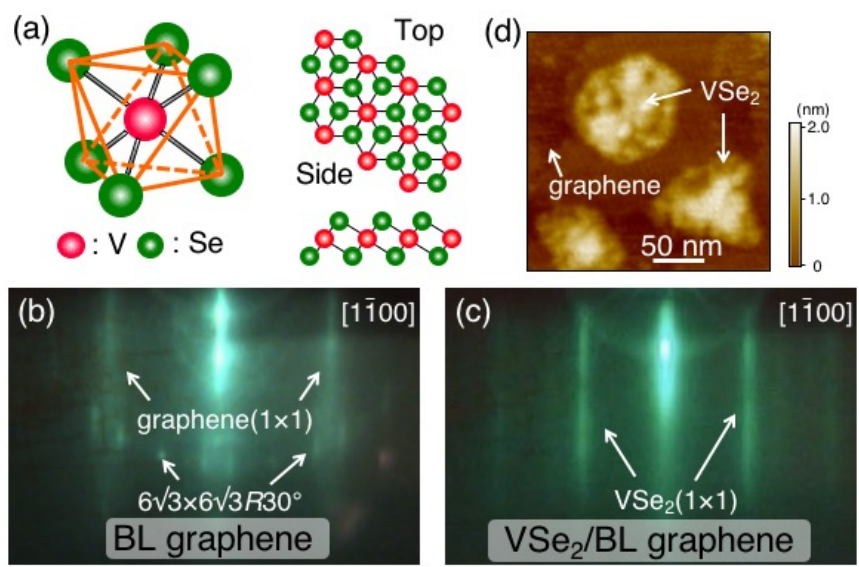

Figure 1 (color online): Y. Umemoto *et al.*

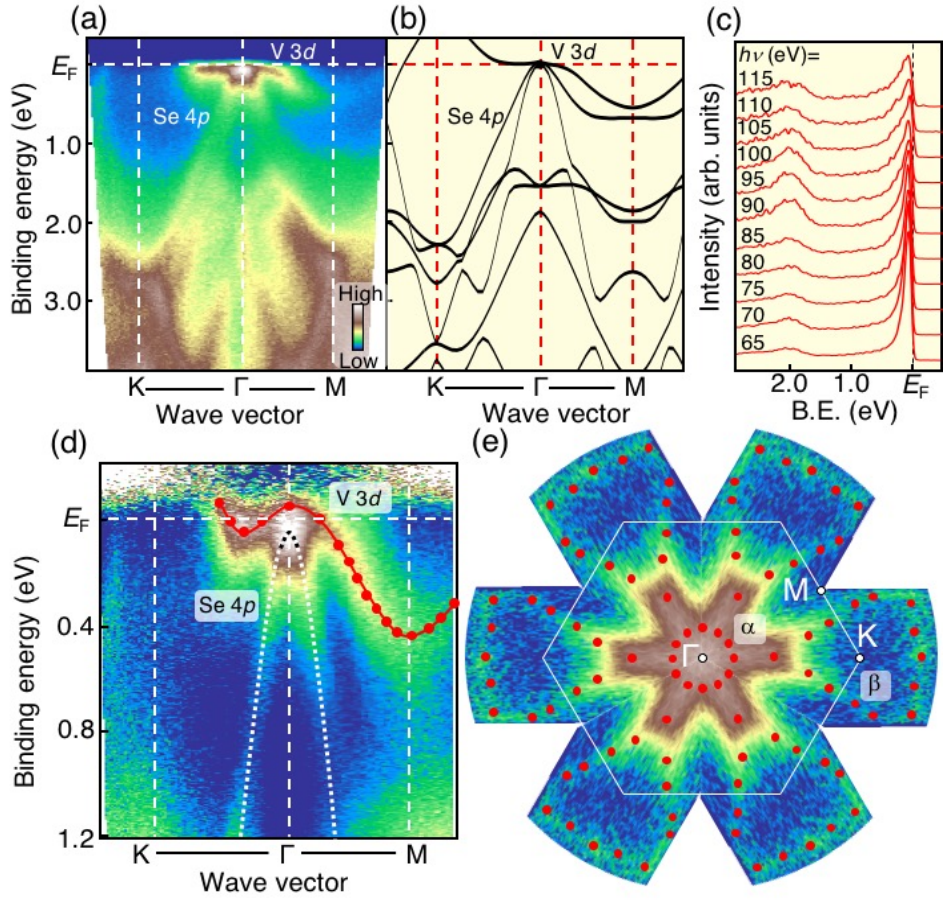

Figure 2 (color online): Y. Umemoto *et al.*

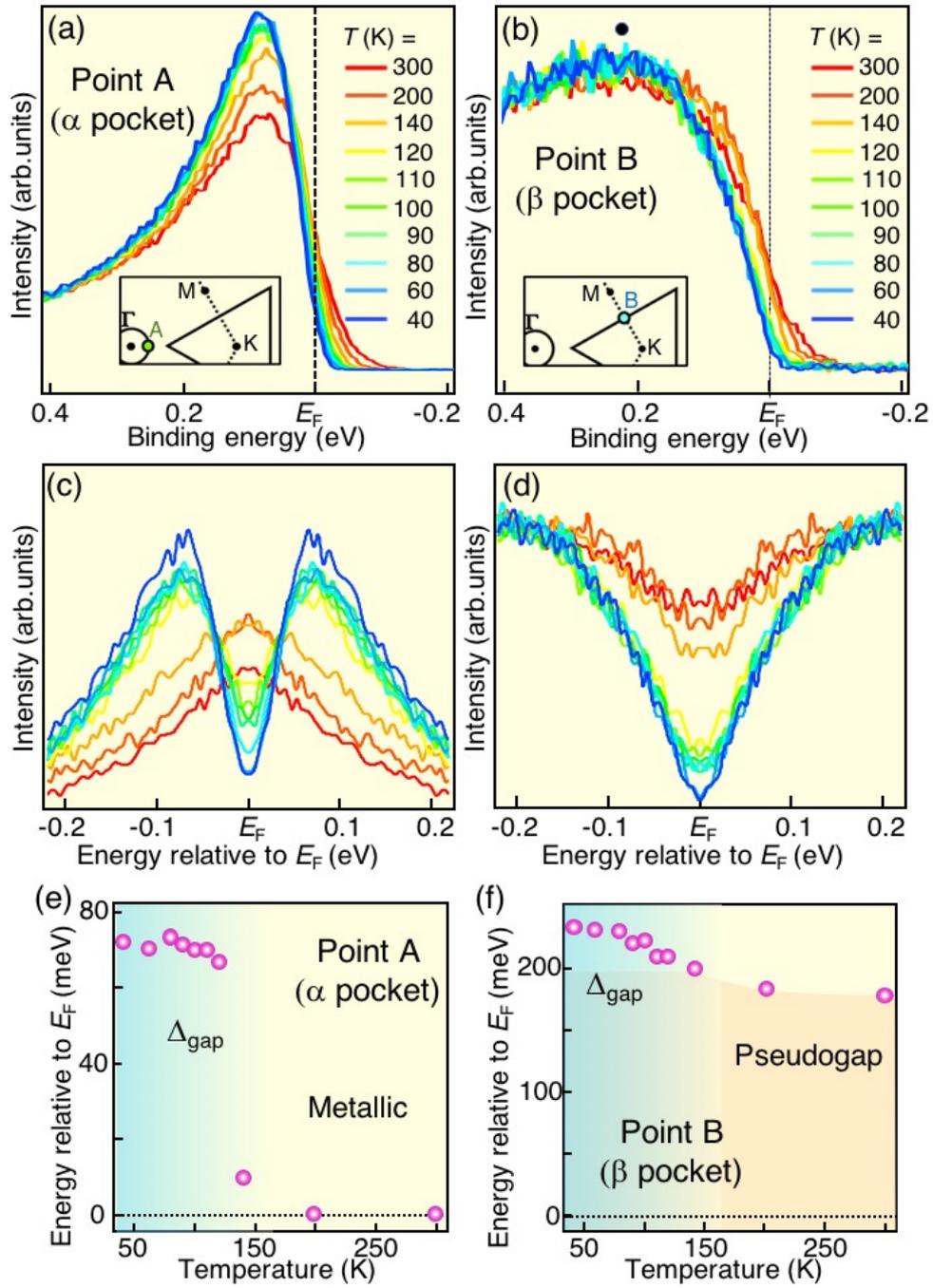

Figure 3 (color online): Y. Umemoto *et al.*

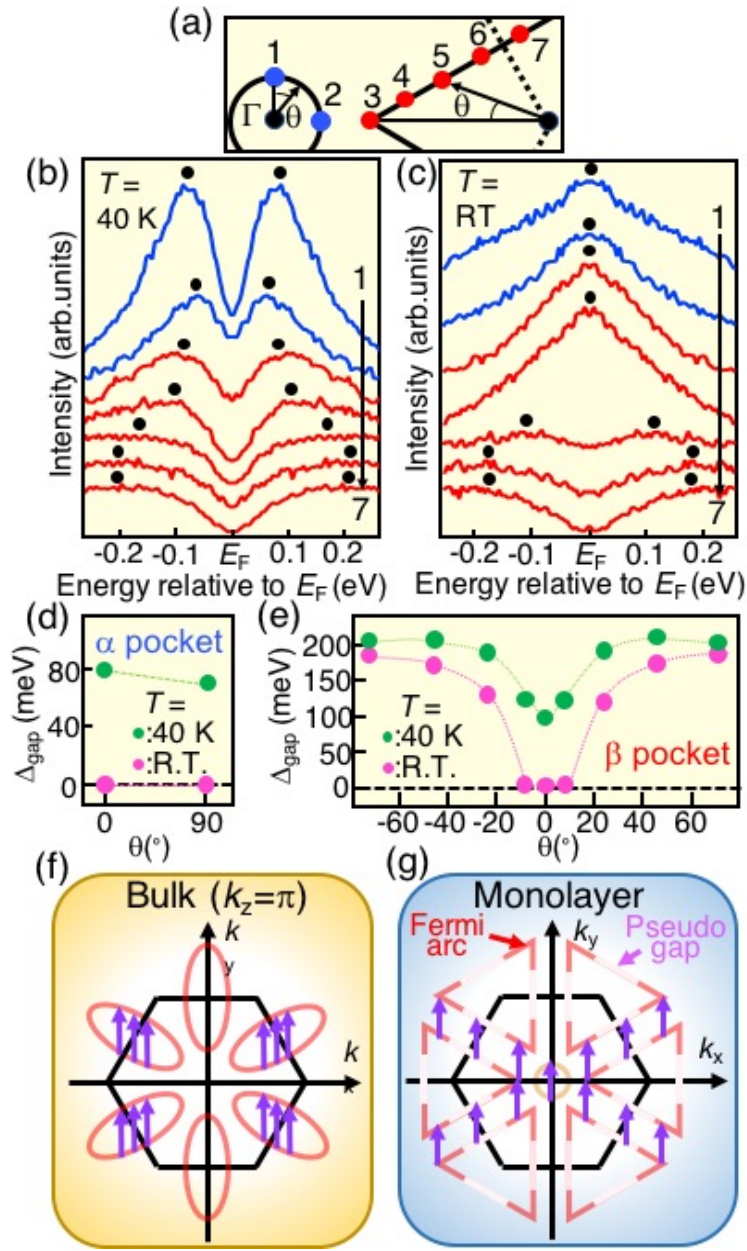

Figure 4 (color online): Y. Umemoto *et al.*